\documentclass[onecolumn,authoryear]{els-mrw} 

\usepackage{amsmath,amssymb,amsfonts,amsthm,makeidx,graphicx}
\usepackage{txfonts}
\usepackage{helvet}
\usepackage{graphicx}
\usepackage{natbib}

\newcommand{\de}{{\rm d}}
\newcommand{\diff}{{\rm d}}
\newcommand{\drr}{\frac{\partial}{\partial r}}

\newcommand{\dr}[1]{\frac{\partial  #1}{\partial r}}

\newcommand{\lp}{ \left(}
\newcommand{\rp}{ \right)}
\newcommand{\lc}{ \left[}
\newcommand{\rc}{ \right]}
\newcommand\aj{{AJ\,}}%
\newcommand\araa{{ARA\&A\,}}%
\newcommand\apj{{ApJ\,}}%
\newcommand\apjl{{ApJ\,}}%
\newcommand\apjs{{ApJS\,}}%
\newcommand\aap{{A\&A\,}}%
\newcommand\mnras{{MNRAS\,}}%
\newcommand\nat{{Nature\,}}%
\newcommand\solphys{Sol.~Phys.\,}%
\begin{document}

\chapter{Mixing processes in stars}\label{chap1}

\author[1]{P. Eggenberger}%

\address[1]{\orgname{University of Geneva}, \orgdiv{Department of Astronomy}, \orgaddress{51. Ch. Pegasi, CH-1290 Versoix, Switzerland}}

\articletag{Chapter Article tagline: update of previous edition,, reprint..}

\maketitle


\begin{glossary}[Nomenclature]
\begin{tabular}{@{}lp{34pc}@{}}
AM &Angular Momentum\\
EoS &Equation of State\\
KH &Kelvin-Helmholtz\\
LTE &Local Thermodynamic Equilibrium\\
PMS &Pre-Main Sequence\\
ZAMS& Zero-Age Main Sequence\\
\end{tabular}
\end{glossary}

\begin{abstract}[Abstract]
Stars play a key role in the evolution of the Universe, as sources of radiation, as dynamical engines, and as chemical factories. Outputs of stellar models are then central to various studies in astrophysics. Stellar physics links fundamental physical aspects to hydrodynamic and magnetohydrodynamic processes, and the validity of stellar models depends directly on the modelling of these complex mechanisms. We describe here the different transport processes at work in stellar interiors and how the modelling of these processes can be improved thanks to the unique ability of asteroseismology, the study of stellar oscillations, to probe the internal structure and dynamics of stars.
\end{abstract}

\begin{keywords}[Keywords]
Stellar physics, Stellar evolution, Stellar processes, Stellar evolutionary models, Stellar rotation, Stellar magnetic fields, Stellar oscillations
\end{keywords}

\begin{glossary}[Key points]
\begin{itemize}
\item Mixing processes in stellar interiors can change the properties of stars and the validity of stellar models depends on the exact modelling of these physical mechanisms.
\item Progressing in the modelling of these complex physical mechanisms requires direct observational information on the properties of stellar interiors that can be provided by asteroseismology, the study of stellar oscillations. 
\item Observations of the internal rotation and of the surface abundances of stars indicate that transport processes are needed for both angular momentum and chemical elements in stellar radiative zones.
\item An important goal of stellar modelling is to obtain a coherent physical description of the various transport processes able to correctly reproduce the observational constraints available for both the transport of angular momentum and chemicals in stellar interiors.
\end{itemize}
\end{glossary}

\section{Introduction: standard stellar models}
\label{secintro}

To determine a theoretical model of a star, the four basic equations of the stellar internal structure have to be solved. These equations express the conservation of different physical quantities. In standard stellar models, the hypothesis of spherical symmetry is used and the star can then be seen as a sphere of hot gas with physical properties that depends on the radius $r$ defined here as the distance to the stellar center. 

The first equation of the internal structure of a star directly follows from the conservation of the mass. The second equation is related to the equation of motion. In a standard stellar model, this equation simply corresponds to the equation of hydrostatic equilibrium, which expresses that, in the interior of a star at equilibrium, all forces acting on a fluid element must compensate each other. The gradient in pressure must then compensates the gravitational acceleration. Interestingly, in the particular case of an equation of state (EoS) with the pressure that only depends on the density (as for example in the case of a degenerate gas), these equations are sufficient to determine the internal stellar structure, namely the values of the pressure, density and the mass interior to the radius $r$ (denoted $M_r$) for any value of the radius $r$. However, the temperature generally also intervenes in the EoS of stellar matter (as in the case of the EoS for a perfect gas). This link between the pressure and the temperature implies that the mechanical equilibrium of the star cannot be separated from the thermal part of the problem. This means that the mechanical and the thermal state of the stellar matter must be solved simultaneously to determine the physical properties of the interiors of stars. 

The third equation of stellar structure expresses the conservation of energy. Two main sources of energy are available for a star. A first possibility consists in extracting energy from the gravitational potential of the star through a slow contraction. The typical lifetime of a star based on this sole source of energy can be simply estimated by dividing the total gravitational energy of the star by its luminosity: this defines the Kelvin-Helmholtz (KH) timescale, which is of the order of 30 - 40 million years in the solar case. A second source of energy is related to thermonuclear reactions once sufficiently high values of temperatures and densities are reached in central stellar layers. The main reaction is hydrogen burning with energy being produced by the transformation of four hydrogen nuclei into one helium nucleus. The typical duration associated to nuclear reactions is much longer than the one related to gravitational contraction with a typical nuclear timescale of about 10 billion years in the solar case.

The last basic equation describing the internal structure of a star expresses the transport of energy from the central stellar layers to the surface. This transport of energy is primarily done by photons; this is the radiative transfer. A photon emitted in the interior of a star is not able to travel on a long distance before interacting with the stellar matter. This characteristic distance between two interactions, the mean free path of the photons, is indeed orders of magnitudes lower than the radius of the star, with typical values between about 0.01 and 1 cm in stellar interiors. Interestingly, gradients in temperature are found to be quite low in stellar interiors. This can be evaluated using the solar values for the radius, central and surface temperatures: an average temperature gradient of about 10$^{-4}$ K cm$^{-1}$ is then obtained. Comparing this value to the mean free path of photons, one concludes that the temperature can be considered as being approximately constant over a photon mean free path. In this case, the photon will hardly see any change in temperature during its travel between two interactions, so that the thermal equilibrium hypothesis can be considered as being locally verified. This is a key hypothesis in the context of radiative transfer, which is the assumption of local thermodynamic equilibrium (LTE). One can then assume that the radiation can be locally described by Planck’s law to obtain the equation of radiative transfer that directly relates the radiative flux to the thermal gradient. The radiative conductivity appearing in this equation for the transport of energy is inversely proportional to the opacity of the stellar matter. The transport of energy through radiative transfer is present in the whole interior of a star. However, in stellar layers that are for instance characterized by high values of opacity, the efficiency of energy transport by radiative transfer could be found to be insufficient to evacuate the heat excess. An additional mechanism for energy transport is then required so that convection takes place in these layers. In addition to its important role for the transport of energy, convection plays of course a fundamental role for the mixing of chemicals (see Sect.~\ref{secconv}). 

A standard stellar model is obtained by solving these four basic equations of the internal structure of a star. To this aim, the microscopic properties of the stellar matter have to be known. An EoS of the stellar matter has first to be used to make the link between the pressure, the temperature and the density for a fixed chemical composition, and to determine the different thermodynamic quantities needed. Concerning the production and transport of energy, a knowledge of the nuclear reaction rates is needed together with the knowledge of the opacities of the stellar matter to determine the radiative transport of energy. Moreover, a criterion for convective instability has to be adopted to determine the location of convective layers in stellar interiors together with a theoretical formalism for the treatment of the effects of convection in these regions. Finally, the evolution of a star can be followed by solving the equations describing the changes in the chemical composition of the stellar matter related to nuclear reactions and the different processes for the transport of chemical elements in its interior as a function of time. These different transport processes are briefly described in the sections below.

\section{Mixing by convection}
\label{secconv}

Within the framework of standard stellar models, radiative zones are assumed to be stable without any global physical processes for the transport of chemical elements. Mixing of chemicals is then assumed to take place solely in convective layers through the efficient turbulent transport by convection (turbulent transport refers to a transport of angular momentum and/or chemicals that is due to the fluid motions). Such a process usually occurs on very short timescales compared to the characteristic timescales of stellar evolution, so that a very efficient mixing leading to an almost instantaneous homogenization of convective regions can be usually assumed in stellar interiors. The determination of a criterion for convective instability is then of prime relevance for chemical mixing in stars.

To obtain such a criterion, one can simply consider a small perturbation of a fluid element around its equilibrium position in the interior of a star. The resulting motion (neglecting the impact of rotation and viscosity) is
\begin{equation}
\rho_{\rm intern} {\de^2 r \over \de t^2} + g(\rho_{\rm intern}-\rho_{\rm extern})=0
\end{equation}
\begin{equation}
\rho_{\rm intern}(r) \approx \rho_{\rm intern}(r_0)+{\de \rho_{\rm intern} \over \de r}(r-r_0)
\end{equation}
\begin{equation}
\rho_{\rm extern}(r) \approx \rho_{\rm extern}(r_0)+{\de \rho_{\rm extern} \over \de r}(r-r_0)  \, ,
\end{equation}
\noindent 
with $r_0$ the radius at the equilibrium position, $\rho_{\mathrm{intern}}$ the density inside the cell, $\rho_{\mathrm{extern}}$ the density characterizing the surrounding stellar medium, and $g$ the local gravity. Using the fact that the external and internal densities are the same at the equilibrium position (i.e $\rho_{\rm intern}(r_0)=\rho_{\rm extern}(r_0)$), one obtains: 
\begin{equation}
\rho_{\rm intern} {\de^2 r \over \de t^2} + g\left ({\de \rho_{\rm intern}\over \de r}
-{\de \rho_{\rm extern} \over \de r}\right )(r-r_0)\approx 0  \, ,
\end{equation} 
with the solution corresponding to an harmonic motion without friction:
\begin{equation}
r-r_0 \sim \exp(i N_{\mathrm{BV}} t) \, .
\label{sol_conv} 
\end{equation}
$N_{\mathrm{BV}}$ is the Brunt-V\"ais\"al\"a frequency, which corresponds to the characteristic oscillation frequency of the cell around its equilibrium position; it is given by :
\begin{equation}
N_{\mathrm{BV}}^2 = \frac{g}{\rho} \left(\frac{\mathrm{d} \rho_{\mathrm{intern}}}{\mathrm{d} r} - \frac{\mathrm{d} \rho_{\mathrm{extern}}}{\mathrm{d} r} \right) \, .
\label{brunt} 
\end{equation}
The sign of this expression directly determines the nature of the motion of the cell. A positive value of $N_{\mathrm{BV}}^2$ will then lead to an oscillatory motion according to Eq.~\ref{sol_conv}, while a negative value will result in an exponential change in the radius. This enables the definition of the following criterion for convective stability :
\begin{equation}
\label{critconv_off}
{\de \rho_{\rm intern}\over \de r} > {\de \rho_{\rm extern}\over \de r} \, .
\end{equation}
In this case, $N_{\mathrm{BV}}$ is real, the amplitude of the oscillatory motion does not increase, which constitutes a necessary and sufficient condition for stability against convection. 
On the other hand, when
\begin{equation}
\label{critconv_on}
{\de \rho_{\rm intern}\over \de r} < {\de \rho_{\rm extern}\over \de r} \, ,
\end{equation}
the Brunt-V\"ais\"al\"a frequency then corresponds to
\begin{equation}
N_{\mathrm{BV}}=-i\sqrt{{g \over \rho}\left ( {\de \rho_{\rm extern}\over \de r}-{\de \rho_{\rm intern}\over \de r}\right )}\, ,
\end{equation}
with the minus sign expressing the fact that the fluid element is moving away from its equilibrium position. The medium is then convectively unstable and this results in a net transport of energy. The temperature of the rising cell being higher than the temperature in the surrounding medium, the cell will transfer heat to these surrounding layers when it will dissolve. Conversely, a fluid element that is moving towards the stellar centre will have a lower temperature than the surrounding layers leading to a cooling of these layers. In addition to leading to an efficient transport of energy from the interior of a star to its surface, these convective motions lead to a very efficient mixing of chemical elements.

The stability criterion given in Eq.~\ref{critconv_off} is expressed as a function of density gradients. This criterion can also be expressed as a function of temperature gradients by making use of the EoS that can be expressed in its general form by:
\begin{equation}
\frac{\mathrm{d} \rho}{\rho} = \alpha \frac{\mathrm{d} P}{P} - \delta \frac{\mathrm{d} T}{T} + \varphi \frac{\mathrm{d} \mu}{\mu} \, ,
\end{equation}
with $P$ the pressure of the gas, $T$ its temperature, $\mu$ its mean molecular weight, which is defined as the mean mass expressed in atomic unit per free particle of the gas, $\alpha = \left( \frac{\partial \ln \rho}{\partial \ln P} \right)_{T,\mu}$, $\delta = - \left( \frac{\partial \ln \rho}{\partial \ln T} \right)_{P,\mu}$, and $\varphi = \left( \frac{\partial \ln \rho}{\partial \ln \mu} \right)_{P,T}$.
We can then expressed the density gradients as:
\begin{equation}\label{drho_ext}
\frac{\mathrm{d} \ln \rho_{\rm extern}}{\mathrm{d} r} = \alpha \frac{\mathrm{d} \ln P_{\rm extern}}{\mathrm{d} r} - \delta \frac{\mathrm{d} \ln T_{\rm extern}}{\mathrm{d} r} + \varphi \frac{\mathrm{d} \ln \mu_{\rm extern}}{\mathrm{d} r} \, ,
\end{equation}
\begin{equation}\label{drho_int}
\frac{\mathrm{d} \ln \rho_{\rm intern}}{\mathrm{d} r} = \alpha \frac{\mathrm{d} \ln P_{\rm intern}}{\mathrm{d} r} - \delta \frac{\mathrm{d} \ln T_{\rm intern}}{\mathrm{d} r} . 
\end{equation}
The absence of the term corresponding to the gradient of mean molecular weight in this expression for the change in the internal density of the fluid element is due to the assumption that the chemical composition of the cell does not change during its motion. The temperature gradients $\nabla$ are then introduced by defining 
\begin{equation}\label{Nabla}
\nabla \equiv  {\mathrm{d} \ln T_{\rm extern} \over \mathrm{d} \ln P} \, ,
\nabla_{\rm intern} \equiv  {\mathrm{d} \ln T_{\rm intern} \over \mathrm{d} \ln P} \, ,
\nabla_\mu \equiv  {\mathrm{d} \ln \mu_{\rm extern} \over \mathrm{d} \ln P} \, .
\end{equation}
Finally, we assume that the motion of the fluid element is sufficiently slow for this cell to be in pressure equilibrium with its surrounding medium; this is equivalent to assuming that the motion of the cell is subsonic. The pressure terms can then be neglected to obtain the Ledoux criterion \citep{led47} for convective stability:
\begin{equation}\label{Ledoux}
\nabla <  \nabla_{\mathrm{intern}} + \frac{\varphi}{\delta}\nabla_{\mu}\, .
\end{equation}
When the stabilizing effects of the $\mu$ gradients are neglected, this criterion corresponds to the Schwarzschild criterion \citep{sch58} for convective stability:
\begin{equation}\label{Sch}
\nabla <  \nabla_{\mathrm{intern}} \, .
\end{equation}

These criteria (Eqs.~\ref{Ledoux} and \ref{Sch}) are used to determine whether a region in the stellar interior is convective or radiative. These criteria for convective stability are based on the balance between forces and determine the boundaries where the total acceleration of the fluid elements vanishes, rather than the velocities. The difference between the boundary where the acceleration of the cell is found to be equal to zero according to the convective criteria and the location where its velocity is equal to zero defines what is usually called the distance of overshooting. In this context, the size of a stellar convective core is determined by adding an overshooting distance denoted $d_{\mathrm{ov}}$ to the location determined by the criteria for convective stability. Such an overshooting distance is usually very simply introduced in stellar evolution codes as an arbitrary extension beyond the boundary defined by the convective criteria that is equal to a given fraction of the pressure scale height $H_P$ defined as $H_P \equiv - \frac{\mathrm{d} r}{\mathrm{d} P} P$:
\begin{equation}\label{dover}
d_{\mathrm{ov}}= \alpha_{\mathrm{ov}} H_P \, .
\end{equation}

\begin{figure}[htb!]
\centerline{\resizebox{0.5\columnwidth}{!}{\includegraphics{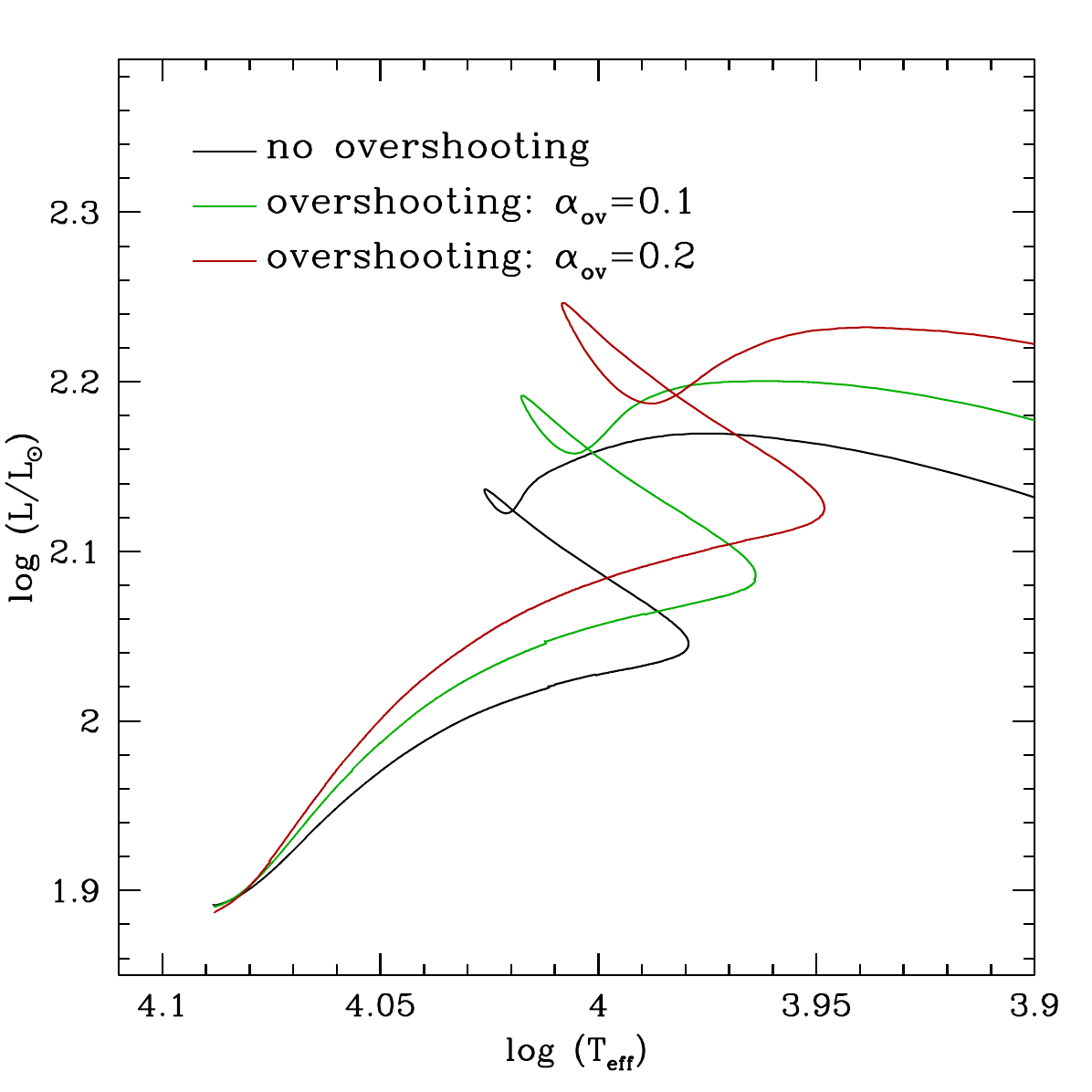}}}
\caption{Main-sequence evolution of 3\,M$_\odot$ stars for different values of the overshooting parameter of the convective core. The black line corresponds to a model without overshooting, while the green and red lines indicate models computed with an overshooting parameter $\alpha_{\mathrm{ov}}$ as defined in Eq.~\ref{dover} of 0.1 and 0.2, respectively. These models are computed for a solar chemical composition and do not account for the effects of rotation.
}
\label{dhrov}       
\end{figure}

\begin{figure}[htb!]
\centerline{\resizebox{0.5\columnwidth}{!}{\includegraphics{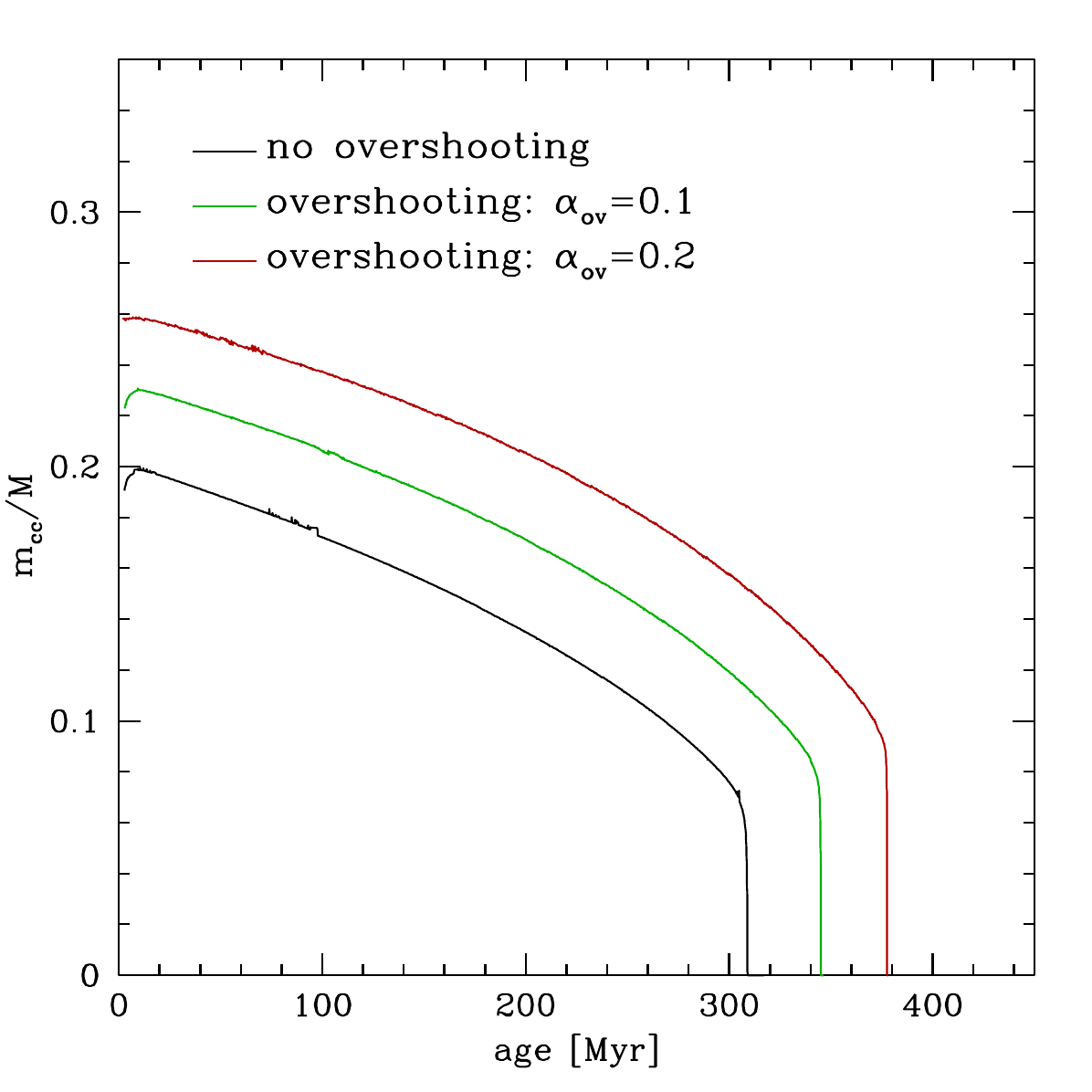}}}
\caption{Evolution during the main sequence of the ratio of the mass of the convective core (m$_{\rm cc}$) to the total mass of the star (M) for the same 3\,M$_\odot$ models shown in Fig.~\ref{dhrov}.}
\label{mccov}      
\end{figure}

The exact location of these convective zones, and in particular the size of convective cores of intermediate-mass and massive stars, constitutes one of the major uncertainties for the computation of stellar models. The impact of overshooting on the effective temperature and luminosity of 3\,M$_\odot$ models during their evolution on the main sequence is illustrated in Fig.~\ref{dhrov}. An increase in the overshooting parameter is shown to result in an increase of the luminosity of the star and to a widening of the main sequence. Both the extension of the main sequence to lower effective temperatures and the increase in luminosity can be related to the increase in the mass of the convective core when the value of the overshooting parameter increases. This is illustrated in Fig.~\ref{mccov}, which shows the evolution of the ratio of the mass of the convective core to the total mass of the star for the same models plotted in Fig.~\ref{dhrov}.
Figure~\ref{mccov} also shows that increasing the overshooting parameter leads to an increase in the main-sequence lifetime as a result of the higher amount of hydrogen fuel available due to the efficient chemical mixing in the convective core.

Observations can be used to constrain the arbitrary value used for the overshooting parameter $\alpha_{\mathrm{ov}}$. This can be done by using the observed width of the main sequence \citep[see e.g.][]{mae81} or the modelling of binary stars \citep[see e.g.][]{cla16}. Moreover, asteroseismic observations can probe the physical properties of stellar interiors, which is particularly useful to constrain the location of the boundaries between convective and radiative layers. This enables to constrain the sizes of convective cores and hence the value of the overshooting distance needed at different evolutionary phases of a star (on the main-sequence and during the core-helium burning phase). Thanks to space missions, observations of solar-like oscillations are now available for a large number stars. Asteroseismic studies using solar-like oscillations detected for main-sequence stars have then been used to constrain the overshooting parameter \citep[e.g.][]{sil13,deh16}. Observations of solar-like oscillations are also available for a large number of post-main sequence stars. They can be used to study convective cores based on the modelling of subgiant stars \citep[e.g.][]{deh11,nol21}. These oscillations can also be used in the case of red-giant stars to obtain constraints on the overshooting parameter needed for the modelling of intermediate-mass star during the main sequence \citep[see e.g.][]{mon10}. This is due to the correlation between the value of the period spacing of oscillations in red giants and the mass of the helium core, which is directly related to the size of the convective core and hence to the value of the overshooting parameter needed during the main-sequence evolution of intermediate-mass stars. The sizes of the cores of red giants during the core-helium burning phase can also be probed thanks to the determination of the values of period spacings for stars in the red clump. Indeed, the internal structure of low-mass stars being similar at helium ignition, the asteroseismic properties of these red-clump stars are not sensitive to the uncertainties related to their modelling during the main sequence \citep[e.g.][]{mon10}. 

Asteroseismic observations of main-sequence $\gamma$ Doradus stars and Slowly Pulsating B (SPB) stars are also helpful for the modelling of the central stellar layers. High-order gravity oscillation modes observed in these stars enable to constrain the chemical gradients at the border of the convective core \citep[e.g.][]{mig08,aer21,far24}. Due to this chemical gradient, a sharp variation is observed in the Brunt-V\"ais\"al\"a frequency, which is directly responsible for a change in the period spacing of gravity modes. This can be used to distinguish between models computed with an overshooting included as a simple extension of the size of the core on a given distance and models predicting smoother chemical gradients at the border of the convective core \citep[e.g.][]{deg10,bow20,aer21,ped21}. In addition to observational constraints, important efforts have been made to compute multi-dimensional hydrodynamic simulations in order to better understand the physical processes at work at convective boundaries and to better parametrize the associated mixing \citep[e.g.][]{mea07,hor21,and22,and24}.

\section{Mixing in radiative zones: the role of rotation}
\label{secrot}

As discussed above in Sect.~\ref{secconv}, chemical mixing is only expected to take place within convective zones in the framework of standard stellar models. For massive stars that are characterized by a convective core and a radiative envelope, this implies that no transport of chemical elements is expected in the radiative envelope and thus chemicals cannot be transported form the central layers to the stellar surface. This is in contradiction with the observations of nuclear burning products at the surface of main-sequence massive stars, and in particular with the observations of surface nitrogen enrichments. This is also in contradiction with the depletion of light elements (lithium, boron) that is observed at the surfaces of different kind of stars. These observations thus suggest that transport mechanisms of chemical elements are not restricted to convective zones and that other mixing processes have to be taken into account in radiative stellar interiors. One of the main physical candidate for such a mixing is related to the different instabilities induced by accounting for rotational effects during the evolution of a star. In this section, we thus briefly review how the effects of rotation can be included in stellar evolution codes to follow the evolution of a rotating star. The modelling of the transport of angular momentum (AM) in stellar interiors is first presented. Then, the modelling of the chemical mixing associated to this transport of AM is discussed by underlying the need for a coherent theoretical description of the simultaneous transport of both AM and chemical elements in radiative stellar layers.

\subsection{Modelling rotational effects in stellar interiors}
\label{modrot}

We begin with a brief description of the theoretical framework describing the physics of rotation as implemented in numerical models of rotating stars \citep[a detailed discussion of the effects of rotation in stellar evolution can be found in][]{mae09}. When rotation is accounted for, assuming that a star is spherically symmetric as done in standard non-rotating stellar models is not necessarily valid. One can however rely on the assumption of shellular rotation \citep{zah92}, i.e. that the angular velocity $\Omega$ in the interior of a rotating star is approximately constant on an isobar, to include rotational effects in one-dimensional stellar evolution codes. Such an hypothesis can be justified by recalling that, due to stratification in stellar interiors, the turbulence related to the different hydrodynamic instabilities induced by rotation is expected to be strongly anisotropic, with a much stronger turbulence in the horizontal than in the vertical direction. Within this theoretical framework, the different physical quantities in the interior of a rotating star can be expressed as a function of the pressure with a small latitudinal perturbation \citep{zah92}.

\subsubsection{Internal transport of angular momentum}
\label{transam}

To follow the evolution of a rotating star, an additional equation (in addition to the four classical equations of the internal structure of a star introduced in Sect.~\ref{secintro}) has to be solved to describe the internal transport of AM and to obtain thereby the internal rotation profile of the star. This evolution of the internal rotation of the star has to account for the turbulent transport of AM by hydrodynamic instabilities triggered by rotation as well as the transport by meridional currents, which are generated by changes in the internal structure of the star, the thermal imbalance induced by the breaking of the spherical symmetry and the lost of AM at the surface of the star  due to stellar winds. This additional equation describing the internal transport of AM can be written as \citep{zah92}:
\begin{equation}
  \rho \frac{{\rm d}}{{\rm d}t} \left( r^{2}\Omega \right)_{M_r} 
  =  \frac{1}{5r^{2}}\frac{\partial }{\partial r} \left(\rho r^{4}\Omega
  U\right)
  + \frac{1}{r^{2}}\frac{\partial }{\partial r}\left(\rho D r^{4}
  \frac{\partial \Omega}{\partial r} \right) \, , 
\label{transmom}
\end{equation}
where $\rho$ denotes the mean density on the isobar and $r$ is the characteristic radius of the isobar. The right-hand side of this equation contains a first advective term describing the AM transport by meridional currents, where $U$ is the radial term of the vertical component of this meridional circulation velocity \citep{mae98}. The other term that appears on the right-hand side of Eq.~\ref{transmom} corresponds to a diffusive transport of AM with a diffusion coefficient $D$. In rotating stellar models accounting solely for hydrodynamic physical processes, transport by the shear instability and meridional circulation are usually considered as the dominating mechanisms in radiative zones \citep[see for instance][for a discussion of the various hydrodynamic instabilities that can be triggered in stellar interiors]{heg00}. In this case, the diffusion coefficient $D$ is then equal to the diffusion coefficient associated to the shear instability $D_{\rm shear}$ \citep{tal97}. Recalling the hypothesis that the turbulence is strongly anisotropic, with a much stronger turbulence in the horizontal than in the vertical direction, this implies that the diffusion coefficient associated to horizontal turbulence ($D_{\rm h}$) is much larger than the one associated to the vertical transport by the shear instability. The modelling of horizontal turbulence in a rotating star is complex and the expression of the associated diffusion coefficient suffers from many uncertainties. This diffusion coefficient $D_{\rm h}$ can be obtained from the balance between the energy dissipated by the horizontal turbulence and the excess of energy in the differential rotation \citep{mae03}. The amplitude of this horizontal turbulence remains however highly uncertain and can be considered as the free parameter of the formalism of shellular rotation that needs to be calibrated in some manner. The important point is to account in a coherent way for the impact of horizontal turbulence, and of the uncertainty related to the free parameter corresponding to the exact amplitude of the associated diffusion coefficient, on the radial transport of AM. This is done within the framework of shellular rotation through the dependence of both the transport by the shear instability $D_{\rm shear}$ and the meridional circulation velocity $U$ on the horizontal turbulence $D_{\rm h}$. 

For the evolution of a rotating star, the internal transport of AM is thus followed by solving Eq. \ref{transmom}. Four boundary conditions are required to solve this equation. Two conditions are first derived by imposing AM conservation at convective boundaries, and two additional boundary conditions are obtained by imposing the absence of differential rotation at these boundaries.

\subsubsection{Internal transport of chemical elements}
\label{transch}

Once the internal rotational properties have been obtained as described above in Sect.~\ref{transam}, the corresponding transport of chemical elements induced by rotation can be derived. The important point is of course to obtain a coherent theoretical description of the simultaneous transport of both AM and chemical elements. Consequently, the turbulent transport of chemical elements by the shear instability in stellar radiative interiors has to be modelled with the same diffusion coefficient $D_{\rm shear}$ as for the transport of AM. Direct numerical simulations of transport by the shear instability indeed show a very similar transport efficiency for AM and chemicals \citep[e.g.][]{pra16}. As for the transport of AM, meridional circulation is also responsible for an advective transport of chemical elements. However, the impact of horizontal turbulence on this advective transport can be treated in a different way for chemicals compared to AM. The advective transport of AM has indeed to be fully followed without having a possibility to correctly describe it through a diffusive approximation. This is due to the fact that, depending on the sign of the velocity $U$, AM transport by meridional circulation can either increase or decrease the degree of radial differential rotation, while a diffusive approximation can only counteract the creation of radial differential rotation. In the case of chemicals, the situation is different: the nature of the transport by meridional currents is still advective, but the simultaneous action of horizontal turbulence on this transport can then be successfully mimicked globally as a purely diffusive process with a global diffusion coefficient $D_{\rm eff}$ associated to these two processes \citep{cha92}:    

\begin{equation}
D_{\rm eff} = \frac{|rU|^2}{30D_{\rm h}}\,,
\label{Deff}
\end{equation}
where $U$ is the radial component of the meridional circulation velocity in the vertical direction and $D_{\rm h}$ corresponds to the horizontal turbulence (see Sect.~\ref{transam}). Rotational mixing in stellar interiors can then be followed with a purely diffusive equation with a total diffusion coefficient $D=D_{\rm shear}+D_{\rm eff}$, which accounts for both the turbulent transport by the shear instability and by the simultaneous impact of meridional currents and horizontal turbulence.

\subsection{Impact of rotational mixing on stellar evolution}

The impact of rotation on the evolution of stellar models is discussed by using the theoretical framework described above in Sect.~\ref{modrot}. The effects of rotation on chemical abundances can already be seen during the pre-main sequence (PMS) evolution of a star. To illustrate these effects, we will consider the case of a solar-type star. At the beginning of the evolution on the PMS, a disc is still present around the star. This has consequences for the rotational properties of the star, since a strong coupling can exist between the star and its disc, which leads to a disc-locking phase. Due to this strong coupling, the surface angular velocity of the star can be assumed to remain approximately constant. As soon as the disc disappears (after about 3 to 10 Myr depending on the assumed disc lifetime), the surface velocity increases as a result of the global contraction of the star during the PMS. As the disc lifetime (and hence the duration of the disc-locking phase) increases, the star is able to evacuate a larger amount of AM during the PMS and will then reach the zero-age main sequence (ZAMS) with a lower surface velocity. This leads to a direct link between the disc lifetime and the surface velocity of a star on the ZAMS. Moreover, increasing the duration of the disc locking phase leads to a higher degree of radial differential rotation in the stellar radiative interior during the PMS. This is due to the fact that the surface velocity of the star remains constant during the disc locking phase, while the velocity in the central layers increases as a result of the contraction. Concerning the impact of rotation on the global stellar properties, one notes that the inclusion of rotational effects results in slightly lower luminosities and effective temperatures during the PMS evolution of the star. These differences between the PMS evolutionary tracks of rotating and non-rotating stellar models are due to the direct effects of rotation on the structure of the star related to hydrostatic corrections due to the centrifugal force.

Regarding the transport of chemical elements during the PMS, one finds that the rapid evolution during this phase implies that rotational mixing is unable to significantly change the internal structure and the global properties of the star. However, the situation is different for the surface abundances of light elements and in particular for the surface lithium abundance. Indeed, as seen in Sect.~\ref{modrot}, rotational mixing can occur in stellar radiative zones, while no transport of chemicals is predicted in these radiative layers within the classical framework of non-rotating models. This mixing induced by rotation is found to be able to transport lithium from the base of the convective envelope to deeper and hence hotter layers where it can then be destroyed. Consequently, including the impact of rotational mixing leads to more lithium depletion during the PMS of solar-type stars compared to the non-rotating case. Moreover, as discussed above, an increase in the disc lifetime increases the degree of radial differential rotation in the interior of a rotating stellar model. This results in an increase of the efficiency of the transport by the shear instability and hence to a higher depletion of lithium. Rotating models of PMS solar-type stars predict thus a correlation between the duration of the disc-locking phase and the surface lithium abundance. Recalling that a longer disc lifetime also implies a lower rotational velocity at the end of the PMS, this leads to an interesting link between the surface velocity and lithium abundance predicted for rotating models. Stars with higher surface rotation velocities on the ZAMS are indeed predicted to exhibit higher lithium surface abundances than slowly rotating stars  \citep[see][for more details]{egg12_pms}. This correlation between surface velocities and lithium abundances predicted for rotating PMS models seems to be in good agreement with observations in the Pleiades \citep[e.g.][]{sod93}.

After the impact of rotation on the evolution during the PMS, we now briefly discuss the effects of rotation on the evolution of main-sequence stars. The evolution during the PMS proceeds on short timescales (given by the KH timescale) compared to the nuclear timescales characterizing the evolution on the main sequence. This implies that rotational mixing has no important impact on the internal structure and global properties of PMS stars (although it can change the surface abundances of light elements as seen above), while it has more time to change significantly the properties of stars in the core-hydrogen burning phase.

\begin{figure}[htb!]
\centerline{\resizebox{0.5\columnwidth}{!}{\includegraphics{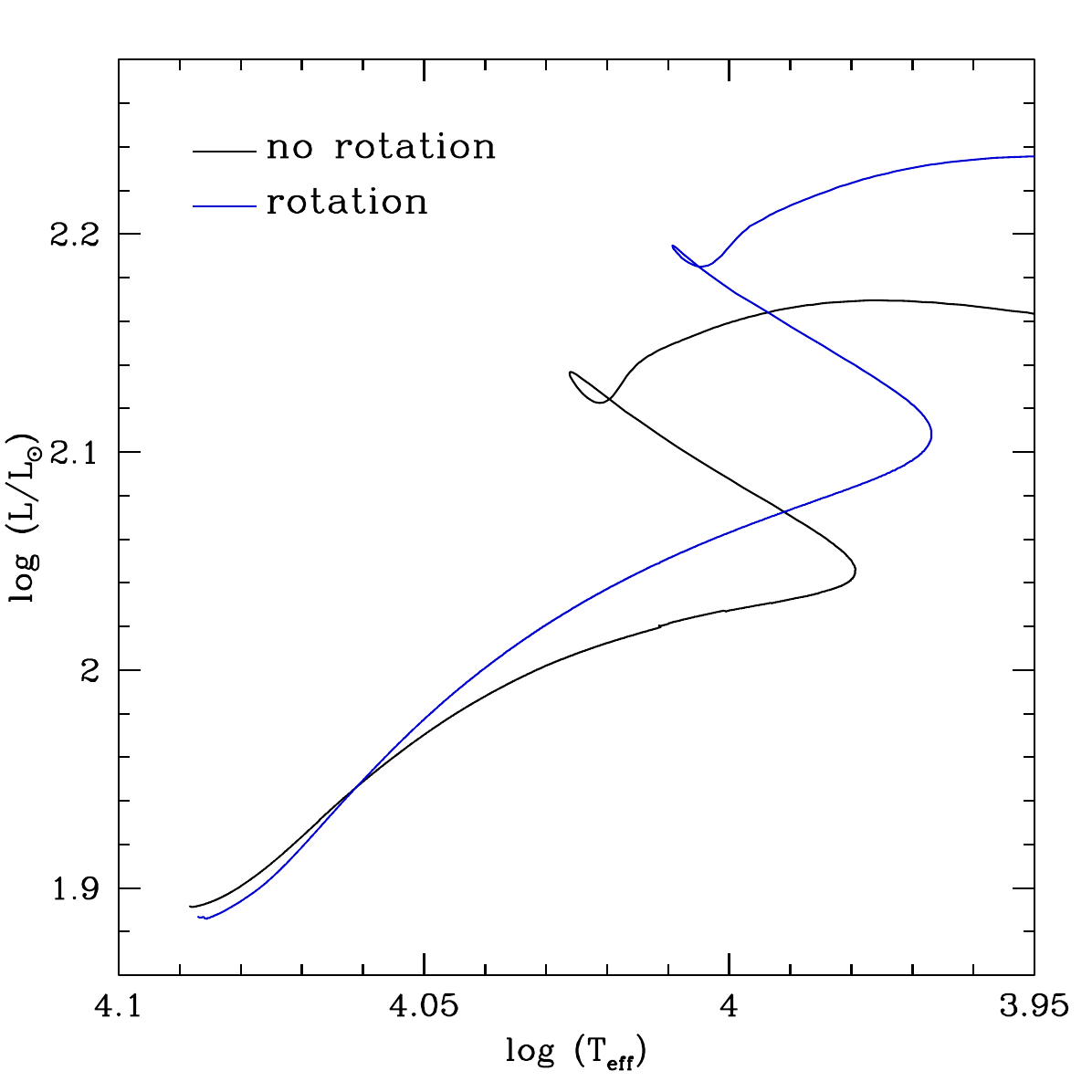}}}
\caption{Main-sequence evolution of 3\,M$_\odot$ stars computed with and without rotational effects. The black line corresponds to a model without rotation, while the blue line denotes a rotating model computed for a rotation velocity on the zero-age main sequence of 150\,km\,s$^{-1}$. Both models are computed for a solar chemical composition and with the same input parameters except for the inclusion of rotation.
}
\label{dhrrot}       
\end{figure}

We first describe the impact of rotation on the evolution of the luminosity and effective temperature during the main sequence. This is illustrated in Fig.~\ref{dhrrot} with two different stellar models of 3\,M$_\odot$: one model is computed without rotation (black line in Fig.~\ref{dhrrot}) and another model is computed within the theoretical framework of shellular rotation presented in Sect.~\ref{modrot} for an initial rotational velocity of 150\,km\,s$^{-1}$ on the ZAMS (blue line in Fig.~\ref{dhrrot}). Both models are computed without the inclusion of overshooting from the convective core discussed in Sect.~\ref{secconv}. Figure~\ref{dhrrot} shows that, at the beginning of the evolution on the main sequence, the evolutionary track corresponding to the rotating model (blue line) is characterized by a slightly lower luminosity and effective temperature compared to the track of the non-rotating model (black line). These differences in the global properties of a rotating model are due to the hydrostatic effects of rotation, with the centrifugal force leading to a decrease in the effective gravity of the rotating model compared to the non-rotating one. This effect is similar to the differences in luminosities and effective temperatures observed above in the case of the PMS evolution of stars with and without rotational effects. However, Fig.~\ref{dhrrot} shows that the lower luminosity of rotating models compared to non-rotating ones is only seen during the first part of the evolution on the main sequence. The increase in the luminosity as evolution proceeds on the main sequence is indeed found to be more rapid for the rotating model than for the non-rotating one. Rotating models exhibit higher luminosities than models without rotation during the second part of the evolution on the main sequence. This is a direct consequence of the transport of chemicals in the radiative interiors of rotating models by hydrodynamic instabilities that is not present in the case of non-rotating models. As discussed in Sect.~\ref{modrot}, this rotational mixing in stellar radiative zones is due to both the transport by the meridional circulation and the shear instability. The contribution of these two processes to the total transport of chemicals depends on the location in the stellar interiors. For stars with a convective core during the main sequence (as the ones shown in Fig.~\ref{dhrrot}), the transport of chemicals is dominated by meridional currents in radiative layers that are close to the border with the convective core. This is due to the presence of strong chemical gradients in these layers, which inhibit the turbulent transport of both chemicals and AM by the shear instability. In radiative layers that are closer to the stellar surface, the situation is different with a transport of chemical elements dominated by the shear instability \citep[e.g.][]{mae00}.

\begin{figure}[htb!]
\centerline{\resizebox{0.5\columnwidth}{!}{\includegraphics{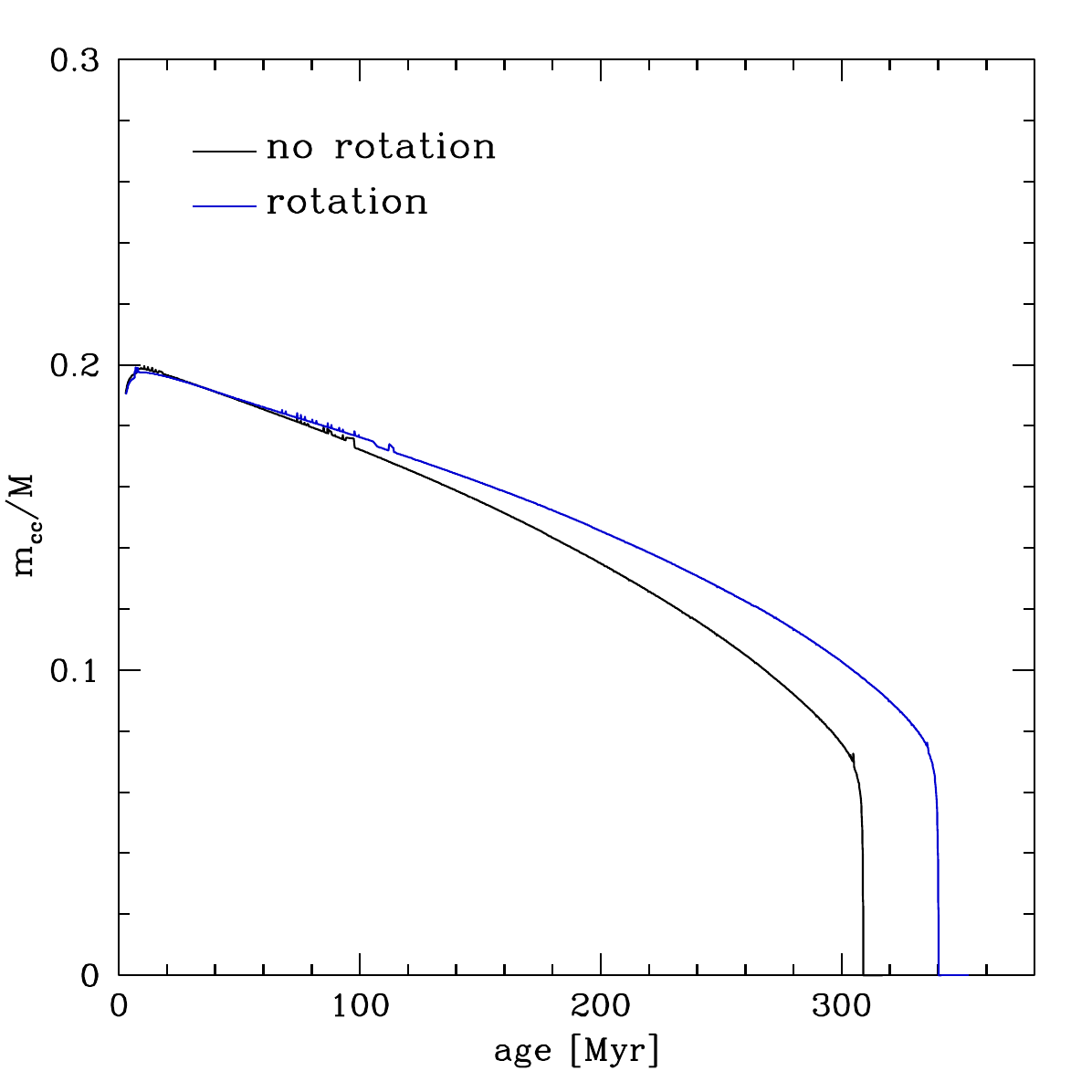}}}
\caption{Evolution during the main sequence of the ratio of the mass of the convective core (m$_{\rm cc}$) to the total mass of the star (M) for the same 3\,M$_\odot$ models with (blue line) and without (black line) rotation shown in Fig.~\ref{dhrrot}.}
\label{mccrot}      
\end{figure}

As shown by the blue line in Fig.~\ref{dhrrot}, the inclusion of rotational effects also leads to a widening of the main sequence. This is similar to the impact of overshooting on the global properties of stellar models discussed in Sect.~\ref{secconv} and shown in Fig.~\ref{dhrov}. As in the case of overshooting, this extension of the main sequence of the rotating model to lower effective temperatures is explained by the increase of the mass of the convective core during the evolution of the star on the main sequence. This is illustrated in Fig.~\ref{mccrot}, which shows the main-sequence evolution of the ratio of the mass of the convective core to the total stellar mass for the same models computed with and without rotation shown in Fig.~\ref{dhrrot}. The increase of the convective core of the rotating model results from rotational mixing that is able to transport fresh hydrogen fuel into the core. As discussed above, such a transport is mainly achieved through meridional currents. Figure~\ref{mccrot} shows that this increase in the mass of the core when rotation is accounted for becomes more and more visible as evolution proceeds, since rotational mixing needs some time before producing significant changes in stellar interiors. As shown in Fig.~\ref{mccrot}, this mixing also leads to an increase of the main-sequence lifetime of the star. Comparing Figs.~\ref{mccov} and \ref{mccrot}, one sees that this increase in the main-sequence lifetime of the rotating model is very similar to the one corresponding to an overshooting parameter of 0.1. Interestingly, this is also the case for the widening of the main sequence due to rotation, as can seen by comparing Figs.~\ref{dhrov} and \ref{dhrrot}. However, these figures show that the increase in luminosity induced by rotational mixing is much larger at the end of the main sequence than the one predicted by the non-rotating model with an overshooting parameter of 0.1. This change in luminosity of the rotating model is indeed found to be very similar to the one predicted by the non-rotating model with an overshoot parameter of 0.2. We thus see that the impact of rotation on the widening of the main sequence and the increase of the main-sequence lifetime is better reproduced by a non-rotating model computed with a lower value for the overshooting parameter of 0.1, while the effects of rotation on the stellar luminosity are better reproduced by a non-rotating model with a larger overshooting parameter of 0.2. This illustrates the fact that rotational mixing not only increases the size of the convective core but also changes the chemical composition in the radiative zone of the star. While both effects contributed to the increase in luminosity of the model computed with rotation, only the increase in the mass of the convective core has a significant impact on the extension of the main sequence to lower effective temperatures \citep[see e.g.][]{egg10_rg}. 
 
We have just seen that the inclusion of the effects of rotation can change the outputs of stellar models. This has interesting consequences in various astrophysical contexts. We have first seen that rotating models can exhibit different effective temperatures and luminosities compared to non-rotating ones (see Fig.~\ref{dhrrot}), with for instance the extreme phenomenon of chemical homogeneous evolution that could result from very efficient rotational mixing \citep[e.g.][]{mae87_hom,dem09,son16,man16}. Transport of chemical elements by rotation induced instabilities in stellar radiative zones could also lead to surface enrichments in nitrogen of B-type stars at solar metallicity and in the Large Magellanic Cloud \citep[e.g.][]{bro11,eks12,egg21}. Similarly, rotation can strongly impact the mass limits between different evolutionary scenarios, as for instance the lower mass limits for single stars evolving into the Wolf-Rayet phase  \citep[e.g.][]{mey05} or undergoing a pair instability supernova \citep[][]{cha12}. The nucleosynthesis of massive stars is also sensitive to the inclusion of rotational effects, as for instance $^{26}$Al at solar metallicity \citep[e.g.][]{pal05} and $^{14}$N at low metallicities \citep[e.g.][]{mey02,lim18}. The pre-supernova stage also depends on the modelling of rotational effects, with an impact on the core collapse, the subsequent phases and the nature and properties of the stellar remnant \citep[e.g.][]{hir04,lim18}. The transport of AM in stellar interiors also appears as a key ingredient when exploring the nature of the progenitors of the long soft Gamma Ray bursts \citep[e.g.][]{hir05,yoo06}, the origin of the primary nitrogen in the early Universe \citep[e.g.][]{chi06}, or the spin and mass limits of stellar black holes and of neutron stars \citep[e.g.][]{heg05}. 

All these changes of the stellar properties induced by rotational mixing depend of course on the exact modelling of the different transport processes at work in stellar interiors. Recalling that this transport of chemicals in stellar interiors is directly connected to the transport of AM through the different instabilities that can be triggered, it is of prime interest to obtain observational constraints on the internal rotation of stars to be able to test the input physics of rotating models. This is now possible thanks to the ability of asteroseismology to probe the internal structure and dynamics of stars.

\subsection{Asteroseismic constraints on the modelling of rotation in stellar interiors}
\label{asterorot}

In this section, we briefly review the observational constraints available to test the modelling of rotational effects included in stellar evolution codes. In particular, the predictions of rotating models of stars computed within the theoretical framework of shellular rotation as described above in Sect.~\ref{modrot} are compared to asteroseismic measurements of the internal rotation of stars.

Helioseismology has enabled to obtain key observational constraints on the internal properties of the Sun. In particular, measurements of frequency splittings of the solar oscillation modes revealed an almost uniform rotation in its radiative zone down to about 0.2 solar radii, while latitudinal differential rotation is present in the convective envelope with a narrow transition between these two regions \citep[see e.g.][]{bro89,els95,kos97,cou03,gar07}. This is an important observational constraint that should be correctly reproduced by rotating stellar models. However, models that account for internal AM transport due to meridional currents and to the shear instability as described in Sect.~\ref{modrot} predict a high degree of radial differential rotation in the solar radiative zone, which is in disagreement with these helioseismic measurements \citep[][]{pin89,cha95,egg05_mag,tur10}. This is a first indication that an efficient AM transport mechanism is required in addition to meridional circulation and the shear instability in stellar radiative interiors.

One can then wonder whether the Sun constitutes a peculiar case or whether an efficient process for internal AM transport is also required for other stars. In the case of main-sequence solar-type stars, it is not easy to constrain radial differential rotation from the sole measurements of the rotational splittings of the detected pressure modes \citep[e.g.][]{lun14}. However, by combining these asteroseismic data with an independent determination of the surface rotation velocity of the star, one can then deduce some constraints on the difference in rotation between the convective envelope and the radiative zone. While being dominated by rotation in the external part of the stellar interior, the mean rotation rate deduced from asteroseismic measurements of rotational splittings of pressure modes retains some sensitivity to the internal rotation of the star. In this context, the presence of radial differential rotation can be revealed through different values for the asteroseismic and surface rotation rates. This kind of analysis has been performed for main-sequence solar-type stars and very small differences have been usually observed between the rotation rate deduced from asteroseismic data and the surface rotation rate \citep[][]{ben15,nie15}. This indicates that an efficient transport mechanism must also be at work in the radiative interior of main-sequence solar-type stars that exhibit a deep convective envelope \citep[][]{bet23}.

Asteroseismology has also enabled to reveal the internal rotation of intermediate-mass and massive main-sequence stars. Numerous and precise determinations of the mean rotation rate of radiative layers that are located at the border of the convective core are now available for gamma Doradus stars thanks to asteroseismic measurements 
\citep[][]{van16,oua17,oua19,li20}. Comparisons with predictions from rotating models indicate that an efficient transport mechanism for AM in radiative layers is also needed for these main-sequence stars in addition to the sole transport by meridional currents and the shear instability \citep[][]{oua19,moy23_gamma}. Interestingly, the rotation rate of the convective core has been recently determined for a small subsample of gamma Doradus stars in addition to the mean rotation rate at the border of the core \citep[][]{oua20,sai21}. These stars are found to rotate nearly uniformly, but with a convective core that rotates slightly faster than the radiative layers close to the border of the core, which confirms the need for an additional transport mechanism \citep[][]{moy24}. Constraints on internal rotation are also available for more massive main-sequence stars from asteroseismology of beta Cephei and slowly pulsating B (SPB) stars. Detailed asteroseismic modelling of a few beta Cephei stars have enabled to reveal their internal rotation, with some stars showing radial differential rotation and others uniform rotation \citep[see e.g.][]{aer03,dup04,bri07,dzi08,sal22,bur23}. Similarly to gamma Doradus stars, near-core rotation rates can be determined from asteroseismology of SPB stars. Preliminary studies suggest that local conservation of AM is not compatible with these measurements \citep[][]{ped22} and it would be interesting to test in more detail the transport mechanisms at work in the interiors of these stars.

In addition to main-sequence stars, the internal rotation of evolved stars has been revealed thanks to the detection of rotational splittings of mixed oscillation modes. Due to their sensitivity to central layers, these oscillation modes are particularly interesting to determine the physical conditions in the core of evolved stars. The first detections of these modes in red giants have shown radial differential rotation with a core rotating more rapidly than the convective envelope \citep[][]{bec12,deh12}. Comparisons with rotating stellar models using the input physics described in Sect.~\ref{modrot} have rapidly shown a large discrepancy of more than two orders of magnitude between predicted and observed core rotation rates. This shows that AM transport by meridional circulation and the shear instability provides an insufficient coupling to explain the internal rotation of red giants, and that an efficient AM transport mechanism is missing in the radiative zone of these stars \citep[][]{egg12_rg,cei13,mar13}. This conclusion has been confirmed with the observations of low core rotation rates for a large number of red giant stars \citep[][]{mos12,dim16,tri17,geh18,tay19}, as well as observations of a low degree of radial differential rotation in subgiants \citep[e.g.][]{deh14,deh20} and in core-helium burning stars \citep[e.g.][]{deh15,mos24}. Based on these determinations of the internal rotation of subgiant and red giant stars, it is possible to constrain the efficiency of the needed additional transport process. It is then found that this efficiency can be precisely determined in the case of evolved stars, and that this determination is insensitive to the rotational history of the star and in particular to the uncertainties related to the modelling of rotation during the main sequence. The AM transport efficiency increases with the mass of the star and decreases as the star evolves during the subgiant phase \citep[][]{egg19}. In the case of red giants, the efficiency of the additional transport process also increases with the mass of the star and when the star ascends the red giant branch \citep[][]{egg17,moy22}. These studies impose fundamental constraints on the missing internal AM transport process that are of prime interest to reveal its physical nature and to thoroughly test different physical candidates for such a transport.

\section{Mixing in radiative zones: the role of internal gravity waves}

As discussed above in Sect.~\ref{asterorot}, asteroseismic measurements indicate that an efficient mechanism for the transport of AM is needed in stellar interiors in addition to meridional currents and transport by the shear instability. One possibility for such a process is the transport of AM by internal gravity waves. These waves, which are generated in convective regions as well as at the boundary between radiative and convective zones, are indeed able to transport energy and AM in radiative layers. They constitute thus an interesting physical candidate for the missing transport mechanism.

\subsection{Transport of angular momentum by internal gravity waves}
\label{igwam}

We begin by briefly discussing the modelling of AM transport related to these waves, and how such a transport can be accounted for in stellar evolution codes within the global framework of shellular rotation introduced in Sect.~\ref{transam} \citep[for greater detail on this modelling of internal gravity waves, see][]{zah97, tal02, tal05, cha05, mae09, mat13}. To account for the additional AM transport by internal gravity waves, the equation describing the internal transport of AM (Eq.~\ref{transmom}) has to be modified by adding a new term on the right-hand side of this equation:

\begin{equation}
\rho \frac{{\rm d}}{{\rm d}t} \left( r^{2}\Omega \right)_{M_r} 
  =  \frac{1}{5r^{2}}\frac{\partial }{\partial r} \left(\rho r^{4}\Omega
  U\right)
  + \frac{1}{ r^2} \drr \left( \rho (D + \nu_{\rm igw}) r^4 \dr{\Omega} \right) 
 - \frac{3}{8\pi} \frac{1}{r^2} \drr{{\cal L}_J} \; ,
\label{transmom_waves}
\end{equation}
where ${\cal L}_J(r)$ corresponds to the local momentum luminosity associated to internal gravity waves. Another diffusion coefficient, $\nu_{\rm igw}$, is also introduced. This coefficient describes the turbulence associated to the existence of shear-layer oscillations, which can be estimated by averaging the diffusion coefficient corresponding to the local shear instability over several shear-layer oscillations \citep[for more detail, see][]{tal05}. The local momentum luminosity associated with gravity waves (${\cal L}_J(r)$) is obtained from the following expression:

\begin{eqnarray}
{\cal L}_J(r) = \sum_{\sigma, \ell, m} {{\cal L}_J}_{\ell, m} \lp r_{\rm conv}\rp
\exp \lc -\tau(r, \sigma, \ell) \rc \; , 
\label{locmomlum}
\end{eqnarray}
where $r_{\rm conv}$ denotes the radius at the transition between the convective envelope and the radiative zone (this formalism corresponds to the transport in low-mass stars characterized by a convective envelope and a radiative interior). With this definition of $r_{\rm conv}$, the quantity ${{\cal L}_J}_{\ell, m} \lp r_{\rm conv} \rp$ thus expresses the AM luminosity at the bottom of the convective envelope. The quantities $\ell$ and $m$ correspond to the degree and the azimuthal order of the wave, and $\sigma$ denotes its local frequency. The local damping rate is expressed by $\tau(r, \sigma, \ell)$, which can be obtained from:

\begin{eqnarray}
\tau = [\ell(\ell+1)]^{3\over2} \int_r^{r_{\rm conv}} 
\lp K + D_{\rm shear} \rp \; {N_{\rm BV} N_{{\rm BV}, T}^2 \over
\sigma^4}  \left({N_{\rm BV}^2 \over N_{\rm BV}^2 - \sigma^2}\right)^{1 \over 2} {\diff r
\over r^3} \; ,
\label{optdepth}
\end{eqnarray}
with $N_{\rm BV}$ the total Brunt-V\"ais\"al\"a frequency as defined in Sect.~\ref{secconv}, and $N_{{\rm BV}, T}$ the thermal part of the total Brunt-V\"ais\"al\"a frequency (defined as $N_{\rm BV}^2 = N_{{\rm BV}, T}^2 + N_{{\rm BV},\mu}^2$, with $N_{{\rm BV},\mu}$ being the chemical composition part of this frequency). $K$ denotes the thermal diffusivity and $D_{\rm shear}$ is the diffusion coefficient of the shear instability as introduced in Sect.~\ref{transam}. The quantity $\sigma$ corresponds to the local Doppler-shifted frequency of the gravity wave. This frequency is related to the wave frequency in the reference frame of the convective envelope $\omega$ by the following relation:

\begin{eqnarray}
\sigma(r) = \omega - m
\lp \Omega(r)-\Omega_{\rm conv} \rp \label{sigma} \; ,
\label{sigma}
\end{eqnarray}
with $\Omega_{\rm conv}$ the angular velocity in the convective envelope (with the assumption of solid-body rotation in convective regions). From Eqs.~\ref{locmomlum} and \ref{optdepth}, one sees that, within this formalism, internal gravity waves will deposit AM in layers where they can be damped by the thermal diffusivity ($K$) and the turbulence induced by the shear instability ($D_{\rm shear}$). With the assumption that the excitation of gravity waves in the convective envelope of the star is the same for prograde and retrograde waves, these equations show that there will be no transport of AM when the angular velocity in the radiative interior is the same as the one in the convective envelope (i.e. in the case of solid-body rotation). Indeed, in this case both fluxes of AM will cancel each other. As soon as a radial gradient of angular velocity is present, the situation changes, since the damping of prograde and retrograde waves will then differ according to Eqs.~\ref{sigma} and \ref{optdepth}. 

This has interesting consequences for the modelling of rotating solar-type stars. Indeed, when this theoretical framework describing the AM transport by internal gravity waves is included in stellar evolution codes in addition to the impact of meridional currents and the shear instability, a transport of AM from the central layers to the surface is observed \citep[e.g.][]{tal05, cha05}. This can be understood by recalling that, for low-mass stars with a deep convective envelope, one expects that the radiative interior rotates faster than the envelope as a result of the surface braking induced by magnetized winds. This implies that prograde internal gravity waves will be damped closer to the base of the convective envelope than retrograde ones. Consequently, AM can be extracted from the central radiative layers of the star thanks to internal gravity waves. This is of particular interest in the context of the rotation profile of the Sun. Rotating models including the transport of AM by internal gravity waves are indeed found to predict a rotation profile in much better agreement with the one deduced from helioseismic data than rotating models that solely account for the transport by meridional circulation and the shear instability \citep[][]{cha05}. However, the wave-like nature of such a transport in the radiative interior of a solar-type star leads to rotation profiles that exhibit weak local gradients in angular velocities that should be detected by helioseismic measurements \citep[e.g.][]{den08,aer19}. Internal gravity waves are thus found to be able to efficiently extract AM from the core of low-mass stars, but another efficient transport process seems to be still needed to reproduce the uniform rotation in the solar radiative zone as deduced from helioseismic measurements. Gravity waves excited in stars with a convective envelope could also impact the rotation profiles of post-main sequence stars. Transport of AM by gravity waves generated by penetrative convection could play an interesting role in explaining the internal rotation of subgiant stars \citep{pin17}. In the case of red giants, AM transport by internal gravity waves is found to be inefficient in their central parts and cannot explain the low core rotation rates obtained from asteroseismic measurements for these stars \citep{ful14,pin17}. Transport of AM by mixed oscillation modes could also change the internal rotation of evolved stars. Contrary to the case of internal gravity waves, this transport by mixed modes is found to be inefficient for stars in the subgiant and the early red giant phase, but could impact the internal rotation of more evolved red giant stars \citep{bel15}. 

Finally, AM transport by internal gravity waves could also change the rotation profile of stars with a convective core. In this case, waves are excited in the convective core and propagate towards the surface in layers of decreasing density, which may lead to an efficient transport of AM. Numerical simulations show that such a transport by internal gravity waves can give rise to different kind of rotation profiles in main-sequence intermediate-mass stars depending on the rotation velocity of the star and on the adopted convective flux for the generation of waves, ranging from uniform rotation, radial differential rotation with a faster rotating core, to retrograde radial differential rotation \citep{rog13,rog15}.

\subsection{Transport of chemicals by internal gravity waves}
\label{igwch}

By changing the rotation profile of a star, internal gravity waves can then impact the efficiency of rotational mixing in stellar interiors. This corresponds to an indirect effect of internal gravity waves on the transport of chemicals in stellar radiative zones. This has been shown to be of particular interest in the context of the surface abundances of light elements for main-sequence low-mass stars. In these stars, accounting for the transport of AM by internal gravity waves leads to a decrease of the degree of radial differential rotation and hence to a decrease of the transport of chemicals through the shear instability \citep[e.g.][]{cha05}. Consequently, rotating models that includes the impact of gravity waves on the internal transport of AM are characterized by higher surface lithium abundances than the corresponding models that only accounts for transport by meridional circulation and the shear instability. When the impact of internal gravity waves on AM transport are taken into account, rotating models are then found to better reproduce the lithium abundances observed at the surface of low-mass stars \citep[see e.g.][]{tal05,cha05}.
 
In addition to the indirect effect of gravity waves on rotational mixing through the change of the rotation of the star, internal gravity waves can have a direct impact on the transport of chemicals in stellar interiors. While at first order and within the linear approximation, gravity waves cannot lead to a direct transport of matter, mixing by waves can occur through different physical mechanisms, such as shear unstable waves \citep[see e.g.][]{pre81,gar91} or diffusion generated by irreversible second order motions \citep[see e.g.][]{sch93,mon94}. This has first been studied in the case of the Sun and low-mass stars. For these stars, accounting for the extra mixing associated to internal gravity waves has been shown to better reproduce the observed surface abundances of light elements \citep[e.g.][]{mon94,mon96}. The direct transport of chemicals by internal gravity waves has also been investigated in the case of intermediate-mass and massive stars on the main sequence. These stars are characterized by a radiative envelope and a convective core that generates gravity waves. It has been shown that the transport of chemical elements in the radiative interiors of these stars can be treated as a diffusive process \citep[][]{rog17}. Further studies have shown that the efficiency of the chemical mixing associated to these waves increases with the stellar mass, while it decreases during the evolution of the star \citep[][]{var23}. It will then be interesting to study in detail the evolution of these rotating intermediate-mass and massive stars by accounting simultaneously for the impact of meridional circulation, shear instability and internal gravity waves on the transport of both AM and chemical elements.

\section{Mixing in radiative zones: the role of magnetic fields}
\label{mod_magn}

Magnetic fields can have an important impact on many aspects of stellar evolution. Here, we only focus on the effects of magnetic fields on the transport of AM and chemical elements in stellar interiors.

\subsection{Transport of angular momentum by magnetic fields}
\label{magch}

Asteroseismic determinations of the internal rotation for stars at different evolutionary stages indicate that at least one efficient AM transport process is missing in stellar radiative zones (see Sect.~\ref{asterorot}). Magnetic fields are prime candidates to ensure an efficient AM transport in stellar interiors and to provide thereby a physical explanation to these asteroseismic observations. 

As a first explanation, large-scale fossil magnetic fields could reproduce the absence of radial differential rotation in stellar radiative zones. This has been invoked to account for the solar rotation profile \citep[e.g.][]{mes87,cha93,rue96,spa10}. A key difficulty is then to be able to simultaneously reproduce the absence of differential rotation in the radiative zone of the Sun and the presence of latitudinal differential rotation in its convective envelope \citep[see][]{gou98,bru06,str11,ace13,woo18}. An increase of the viscosity in the radiative interior of the Sun seems also to be needed in addition to a large-scale fossil magnetic field to correctly reproduce the solar rotation profile as deduced from helioseismic measurements \citep{rue96,spa10}. In the case of post-main sequence stars, large-scale fossil magnetic fields could also ensure uniform rotation in the radiative interiors. Assuming that radial differential rotation is present in the convective envelopes of evolved stars, this could explain the internal rotation of red giant stars \citep{kis15, tak21}. Detailed asteroseismic modellings performed for red giant stars are however found to disfavor a uniform rotation profile in the radiative interior of these stars and suggest that the transition in the rotation rate occurs within the radiative zone \citep[e.g.][]{kli17,fel21}. 
 
The impact of magnetic instabilities on the internal transport of AM is then particularly interesting to consider. In stellar radiative layers, the Tayler instability \citep{tay73} seems to be the first magnetic instability to be triggered \citep{spr99}. The winding-up of an initial weak magnetic field by differential rotation can produce strong toroidal fields that are then subject to this instability. A field amplification cycle that produces a small-scale time-dependent dynamo based on the Tayler instability has been proposed by \cite{spr02}. This process, referred to as the Tayler-Spruit dynamo, has been tested in numerical simulations. It is of course particularly difficult for these simulations to be performed under realistic stellar conditions \citep[see][]{bra17}, but recent efforts indicate that a dynamo process based on the Tayler instability could be potentially at work in stellar radiative zones \citep[see e.g.][]{pet23,bar23}.

To include the effects of magnetic instabilities in stellar evolution codes, Eq.~\ref{transmom} for the transport of AM has to be slightly modified by adding a viscosity associated to the magnetic process:
\begin{equation}
  \rho \frac{{\rm d}}{{\rm d}t} \left( r^{2}\Omega \right)_{M_r} 
  =  \frac{1}{5r^{2}}\frac{\partial }{\partial r} \left(\rho r^{4}\Omega
  U\right)
  + \frac{1}{r^{2}}\frac{\partial }{\partial r}\left(\rho (D + \nu_{\mathrm{magn}}) r^{4}
  \frac{\partial \Omega}{\partial r} \right) \, . 
\label{transmom_magn}
\end{equation} 
The magnetic viscosity $\nu_{\rm magn}$ corresponds to the vertical transport of AM by the Tayler instability, which is given by the following general expression \citep{egg22_magn} that encompasses the original prescription by \cite{spr02} and the revised version of \cite{ful19}: 
 \begin{equation}
 \nu_{\rm magn} =
  \; \frac{\Omega \; r^2}{q} \;
 \left(  C_{\rm T} \; q \; \frac{\Omega}{N_{\rm BV, eff}}  \right)^{3/n} \; 
\left(\frac{\Omega}{N_{\rm BV, eff}}\right) \; .
 \label{nu_generale}
 \end{equation}
In this expression, $q= \left| \frac{\partial \ln \Omega}{\partial \ln r} \right| $ is the shear parameter, $C_{\rm T}$ is a dimensionless calibration parameter introduced to account for uncertainties on the adopted damping timescale of the azimuthal magnetic field, and $N_{\rm BV, eff}$ is the effective Brunt-V\"{a}is\"{a}l\"{a} frequency that takes into account the reduction of the stabilizing effect of the entropy gradient through thermal diffusion:
 \begin{equation}
 N^2_{\rm BV, eff}= \frac{\eta}{K} N_{{\rm BV},T}^2 + N_{{\rm BV}, \mu}^2 \; ,
 \label{Neff}
 \end{equation}
with $\eta$ and $K$ the magnetic and thermal diffusivities, respectively. For $n=1$ (and $C_{\rm T}=1$), the general Eq.~\ref{nu_generale} gives the expression corresponding to the original prescription by \cite{spr02}, and for $n=3$ it corresponds to the revised prescription proposed by \cite{ful19}.
The additional AM transport by the Tayler instability is accounted for through the viscosity $\nu_{\rm magn}$ only when the shear parameter $q$ is larger than a threshold value $q_{\rm min, magn}$ given by:
\begin{equation}
q_{\rm min, magn} = C_{\rm T}^{-1} \left(\frac{N_{\rm BV, eff}}{\Omega}\right)^{(n+2)/2} \left(\frac{\eta}{r^2 \Omega}\right)^{n/4} \; .
\label{qmin_generale}
\end{equation}

Rotating models accounting for both hydrodynamic instabilities and the Tayler instability can be computed based on these equations. The Tayler-Spruit dynamo in its original form ($n=1$ and $C_{\rm T}=1$ in Eqs.~\ref{nu_generale} and \ref{qmin_generale}) is then found to provide an efficient transport of AM in stellar radiative zones. Rotating models of the Sun computed with the Tayler-Spruit dynamo predicts a nearly flat rotation profile in the main part of the radiative zone in agreement with helioseismic measurements \citep{egg05_mag,egg19_sun}. Moreover, the Tayler-Spruit dynamo predicts an even stronger coupling in the horizontal than in the radial direction, which is able to account for the sharp transition between latitudinal differential rotation observed in the solar convective envelope and the absence of latitudinal differential rotation in the radiative interior of the Sun. Models accounting for AM by the Tayler instability are also found to provide an interesting explanation to the internal rotation of main-sequence gamma Doradus stars \citep{moy23_gamma,moy24}.

While the original Tayler-Spruit dynamo provides a physical explanation to the internal rotation of the Sun, it does not enable to correctly reproduce the low core rotation rates of evolved stars \citep{can14,den19}. A revised prescription for this process ($n=3$ and $C_{\rm T}=1$ in Eqs.~\ref{nu_generale} and \ref{qmin_generale}) that predicts a more efficient AM transport and hence lower core rotation rates for subgiant and red giant stars in better global agreement with observed values was then proposed by \cite{ful19}. Similarly, a calibrated version of the original Tayler-Spruit dynamo ($n=1$ and $C_{\rm T}=216$ in Eqs.~\ref{nu_generale} and \ref{qmin_generale}) enables to better reproduce the core rotation rates of red giants \citep{egg22_magn}. However, these revised prescriptions have still difficulties to correctly reproduce the internal rotation of stars at various evolutionary stages \citep[see e.g.][]{egg19_full,den20,moy24}. Another magnetic instability, the azimuthal magneto-rotational instability, could also impact the transport of AM in stellar interiors. Rotating models that includes a simple parametric prescription aiming at describing the AM transport by this azimuthal magneto-rotational instability have shown interesting results for evolved stars \citep{spa16,moy23_amri}.

\subsection{Transport of chemicals by magnetic fields}
\label{magch}

The direct transport of chemical elements by the Tayler instability is found to be almost negligible compare to the transport of angular momentum \citep{spr02,ful19}. This is due to the fact that chemicals are transported by the motions of the fluid, while the efficient transport of AM is ensured by the magnetic stresses. The efficient transport of AM related to magnetic fields has however an important indirect impact on the transport of chemicals in stellar interiors. In magnetic rotating models computed with the Tayler-Spruit dynamo, transport of chemical elements by the shear instability is then strongly reduced. However, rotational mixing associated to meridional circulation is more efficient in these models than in models without magnetic fields. We thus see a competition between these two effects to determine the global impact of magnetic instabilities on rotational mixing. Recalling that the efficiency of the transport by meridional currents increases with the rotation velocity of the star, we conclude that the indirect effects of magnetic instabilities on the transport of chemicals will also depend on this rotation velocity. For instance, solar-type stars are rotating slowly as a result of the efficient braking of their surface by magnetized winds. In this case, only a negligible increase is observed for  the efficiency of the transport of chemical elements by meridional currents, which is not able to compensate for the large decrease in the transport efficiency by the shear instability. Consequently, the inclusion of the impact of magnetic instabilities results in a strong decrease in the efficiency of rotational mixing in the interiors of models of solar-type stars. However, for more massive stars that are faster rotators, the increase in the transport efficiency of chemicals by meridional circulation dominates over the decrease of the transport by the shear instability when the impact of magnetic fields is accounted for. Rotating models of massive stars computed with the Tayler-Spruit dynamo are then found to be more efficiently mixed than models including only transport by meridional circulation and the shear instability. An important challenge is then to obtain rotating models that can reproduce in a consistent and physically motivated way both the observational constraints available for the internal transport of AM and of chemical elements. Preliminary results in this direction have been obtained in the case of the Sun \citep[e.g.][]{egg22_sun} and it will then be interesting to study the impact of these transport processes constrained by asteroseismic measurements on the evolution of more massive stars and for stars in low-metallicity environments. 

\section{Summary}

Mixing processes in stellar interiors can impact all outputs of stellar models. The validity of these models depends then on our understanding of these complex physical mechanisms. Progress in modelling these processes requires direct observational constraints on the physical properties of stellar interiors. Asteroseismology, the study of stellar oscillations, can provide us with key information about the physical properties in central stellar layers and about the internal rotation of stars. The aim, and one of the main current challenge of stellar physics, consists thus in developing a coherent modelling of the various transport processes at work in stellar interiors based on these asteroseismic constraints.

\seealso{\cite{mae09,sal17}}

\bibliographystyle{Harvard}
\bibliography{biblio_els}

\end{document}